\documentstyle[12pt,prb,aps,psfig]{revtex}
\begin{document}
\title{Dynamical Stability of an Ion in a Linear Trap\\ as a Solid-State Problem of Electron Localization\\ \ \\}
\author{G.P. Berman$^1$, A.R. Bishop$^1$,  D.F.V. James$^1$,
R.J. Hughes$^2$, and D.I. Kamenev$^{1,3}$\\ \ \\}
%
\address{$^1$ Theoretical Division, Los Alamos National
Laboratory, Los Alamos NM  87545\\
\noindent
$^2$ Physics Division P-23, MS H-803, Los Alamos National Laboratory, Los Alamos NM  87545\\
\noindent
$^3$ Center for Nonlinear Studies, MS B-258, Los Alamos National Laboratory, \\ Los Alamos NM  87545\\ \ \\}
\maketitle

\vspace{-0.2cm}
\begin{abstract}
When an ion confined in a linear ion trap interacts with a coherent
laser field, the internal degrees of freedom, related to the electron 
transitions, couple to the vibrational degree of freedom of the ion. 
As a result of this interaction, quantum dynamics of the vibrational
degree of freedom becomes complicated, and in some ranges of parameters
even chaotic. 
We analyze the vibrational ion  dynamics using a formal analogy with the
solid-state problem of electron localization. In particular, we show 
how the resonant approximation used in analysis of the ion  dynamics,  
leads to a transition 
from a two-dimensional (2D)
to a one-dimensional problem (1D) of electron localization.
The localization length in the solid-state problem is estimated
in cases of weak and strong interaction between the cites of the
2D cell by using the methods of resonance perturbation theory, common in
analysis of 1D time-dependent dynamical systems.

\end{abstract}
%
\newpage

\section*{Introduction}
The problem of quantum dynamics for Hamiltonian systems with time-periodic
perturbation (TPS) can be formulated in terms of an equivalent  solid-state
problem of electron localization on a lattice. Such kinds of connections
were
discussed  for different models in~\cite{2,3} (see also references therein).
However, most results are obtained for quantum kicked systems,
such as a quantum kicked rotor or a quantum kicked oscillator.
These systems are convenient for both analytical and numerical
analysis because instead of differential equations one can use
discrete quantum maps. At the same time, the existence of periodic
kicks suggests that the external field involves an infinite number
of harmonics with equal amplitudes.
In more common physical situations, there are only few harmonics
in the perturbation. In particular, such a situation occurs when
an ion trapped in a linear ion trap interacts with two laser field
with close frequencies~\cite{1}. In this case, the internal degree
of freedom of the ion (related to the electron dynamics) interacts with
the vibrational degree of freedom.
This interaction can result in complicated and even chaotic dynamics of
the vibrational degree of freedom of the ion. The analysis of the
stability of the ion  in this system can be performed using a model
of a quantum harmonic oscillator perturbed by  a monochromatic wave (MPO)~\cite{1}.

In this paper we show that the problem of stability of a MPO can
be formulated  in terms of localization of an electron in a
2D solid-state system (SSS). The resonance approximation,
common in treatment of TPSs, is used
to reduce the effective dimensionality of the SSS in the case of relatively
small interaction of the ion with the laser field.
In order to compare two completely different systems, 
a similarity in the formal description of the 
TPS and a space-periodic SSS is exploited. 
Namely, in the TPS we use the time-periodicity of the perturbation and
employ a Floquet formalism, while in the SSS we exploit
a space-periodicity and use the Bloch theorem.   
   
The paper is organized as follows. In Section~I we describe the MPO
model of an ion trapped in a linear ion trap and interacting with two
laser fields with close frequencies.
In Section~II, we
discuss the general procedure which allows to treat an 1D
TPS on the same ground as a 2D SSS.
In the case of a small perturbation, the resonance approximation
is used in Section~ III to decrease the effective
dimensionality of the SSS from two to one.
The localization length in the SSS is estimated in Section~IV
by calculating the size of the chaotic region in the
corresponding TPS, in the situation when the interaction between the sites
of 2D SSS is strong. Concluding remarks are given in Conclusion.


\section{The Vibrational Hamiltonian}
In the following~\cite{1}, we assume that two laser beams, designated the
pump (p) and the Stokes (s), with slightly different frequencies,
$\omega_p$ and $\omega_s$, respectively, interact with the
ion trapped in a linear ion trap.
Both beams are assumed to be plane polarized in the
z-direction with the amplitudes of the electric field, ${\cal E}_z^{(p)}$
and ${\cal E}_z^{(s)}$, and the wave vectors, ${\bf k}_{p}$ and ${\bf k}_{s}$.
The Hamiltonian, including
the effect of the harmonic evolution of the ion along the weak
axis of the trap (but excluding the internal free evolution)
is,
\begin{equation}
\label{3}
\hat{\cal H}=
{\hat{p}^2\over{2M}}+{{M\omega^2\hat{x}^2}\over{2}}+{{\varepsilon}\over{k}}
\cos(k\hat x-\Omega t),
\end{equation}
where where $\hat p$ and $\hat x$ are the $x$-components of the momentum
and the coordinate of the ion, $t$ is the time,
$M$ is the mass of the ion, $\omega$
is the frequency of the ion vibrations in the linear trap,
$\varepsilon=2\chi k \left|{\cal E}^{(p)}_{z}{\cal E}^{(s)\ast}_{z}\right|$,
$\Omega=\omega_ {p}-\omega_ {s}$,
$k=\left({\bf k}_{p}-{\bf k}_{s}\right)\cdot\bf{e}_{x}$,
$\bf{e}_x$ is a unit vector in $x$-direction,
$\chi= A\pi \epsilon_{0}/4 \nu^{3}\Delta$
($\nu$ and $A$ being,
respectively, the wavenumber and the Einstein
A coefficient for the transition between the upper and lower
manifolds, $\Delta$ is the laser detuning and $\epsilon_{0}$
the permitivity of free space).

In the dimensionless form the Hamiltonian (\ref{3}) reads,
\begin{equation}
\label{6}
\hat H={\hat{\cal H}\over (M\omega^2/k^2)}=
-{{h^2}\over{2}}{{\partial^2}\over{\partial X^2}}+
{{X^2}\over{2}}+\epsilon\cos(X-\mu\tau)=\hat H_0+V(X,\tau),
\end{equation}
where $\hat H_0$ is the Hamiltonian of a linear oscillator,
\begin{equation}
\label{1}
X=kx,\qquad \tau=\omega t,\qquad
\epsilon={{\varepsilon k}\over{M\omega^2}},\quad h={{\hbar
k^2}\over{M\omega}},\quad \mu={{\Omega}\over{\omega}}=
N+\delta.
\end{equation}
Here $h$ is a dimensionless Planck constant, $N$ is the
(positive integer) resonance number, and
$\delta$ is the detuning from the resonance.

The classical analog of the Hamiltonian (\ref{6}) is,
\begin{equation}
\label{H_cl1}
H={{X^2}\over{2}}+{{P^2}\over{2}}+
\epsilon\cos(X-\mu\tau),
\end{equation}
where $P=kp/M\omega$ is the dimensionless momentum.

In the action-angle variables, ($I,\vartheta$), the classical Hamiltonian
(\ref{H_cl1}) takes the form,
\begin{equation}
\label{H_cl}
H=I+
\epsilon\cos[kr(I)\sin\vartheta-\mu\tau],
\end{equation}
where $X=kr(I)\sin\vartheta$, $P=kr(I)\cos\vartheta$,
$kr=\sqrt{X^2+P^2}=\sqrt{2I}$ is the dimensionless amplitude of oscillations,
$I$ is the dimensionless action, measured in units of
$I_0=M\omega/k^2$, and $\vartheta$ is the phase of oscillations.
%
%
\section{Connection with a 2D solid-state Localization Problem}

We write the solution to the Schr\"odinger equation,
\begin{equation}
\label{7}
ih{{\partial\Psi(X,\tau)}\over{\partial\tau}}=\hat H\Psi(X,\tau),
\end{equation}
in the form of series over the eigenfunctions,
$|n\rangle\equiv\phi_n(X)$, of the harmonic oscillator
Hamiltonian, $\hat H_0$,
\begin{equation}
\label{10}
\Psi(X,\tau)=\sum_{n=0}^\infty c_n(\tau)|n\rangle.
\end{equation}
Then we obtain the equations for the complex amplitudes, $c_n(\tau)$,
\begin{equation}
\label{11}
ih{{dc_m(\tau)}\over{d\tau}}=h(m+1/2)c_m(\tau)+\epsilon
\sum_{n=-m}^\infty\langle m|\cos(X-\mu\tau)|m+n\rangle c_{m+n}(\tau)=
\end{equation}
$$
 h(m+1/2)c_m(\tau)+{{\epsilon}\over{2}}\sum_{n=-m}^\infty
 \Bigg(e^{-i\mu\tau}F_{m,m+n}+e^{i\mu\tau}F^*_{m,m+n}\Bigg)c_{m+n}(\tau).
$$
In Eq.~(\ref{11}), $F_{m,m+n}$ is the matrix element~\cite{Gradstein},
\begin{equation}
\label{12}
F_{m,m+n}=\langle m|e^{iX}|m+n\rangle=
{i^nh^{n/2}e^{-h/4}\over2^{n/2}\sqrt{(m+1)(m+2)\dots(m+n)}}
L_m^n\left(\frac{h}2\right),
\end{equation}
where $L_m^n$ is the Laguerre polynomial.
When $m\gg 1$,
the Laguerre polynomials can be
expressed in terms of the Bessel functions, $J_n$~\cite{Gradstein}, as
\begin{equation}
\label{Laguerre}
L_m^{n}\left(\frac h2\right)=\left(\frac{2m}h\right)^{n/2}J_{n}(\sqrt{2mh}),
\end{equation}
where the argument of the Bessel function,
$\sqrt{2mh}=kr_m$, is the quantized
dimensionless amplitude of oscillations of the harmonic oscillator.
Using Eqs.~(\ref{12})~and~(\ref{Laguerre})
the matrix elements can be written in the form,
\begin{equation}
\label{m_el}
F_{m,m+n}=\frac{i^n m^{n/2} e^{-\frac{h}4}}
{\sqrt{(m+1)\dots (m+n)}}J_{n}(\sqrt{2mh}).
\end{equation}

Since the Hamiltonian (\ref{6}) is periodic in time
the solution of the Schr\"odinger equation (\ref{11})
can be written as,
\begin{equation}
\label{13}
c^{q}_m(\tau)=e^{-i\sigma_q\tau/h}A^{q}_m(\tau),
\end{equation}
where $\sigma_q$ is a quasienergy, measured in units of $M\omega^2/k^2$,
$c^{q}_m(\tau)$ is the
quasienergy (QE) function, and $A^{q}_m(\tau)$ is a 
periodic function with the period, $T=2\pi/\mu$,
\begin{equation}
\label{14}
 A^{q}_m(\tau+2\pi/\mu)=A^{q}_m(\tau).
\end{equation}
The quasienergy functions are the eigenfunctions of the 
evolution operator, $\hat U$, for one period, $T$, of the external field,
\begin{equation}
\label{U}
\hat U(T)c^{q}_m(\tau)=e^{-i\sigma_qT/h}c^{q}_m(\tau).  
\end{equation}
The evolution operator for one period of the external field, $\hat U(T)$ is,
\begin{equation}
\label{U_H}
\hat U(T)=\hat Te^{-i\int_0^T\hat H(\tau)d\tau},
\end{equation}
where $\hat T$ is the ordering operator,
$\hat H(\tau)$ in our problem is given by Eq.~(\ref{6}).
In our numerical calculations, presented below, we consider 
only the QE states at the time $\tau=0$, so that
$c_m^q(0)=A_m^q(0)\equiv A_m^q$.\footnote{If the 
spectrum, $\sigma_q$, and QE functions, $A_m^q$, 
are known, one can trace evolution of the quantum system 
at the times $\tau_s=sT$, where $s=0,\,1,\,2\dots$
(see, for example \cite{31}).}   

 \begin{figure}[tb]
 \begin{center}
 \mbox{\psfig{file=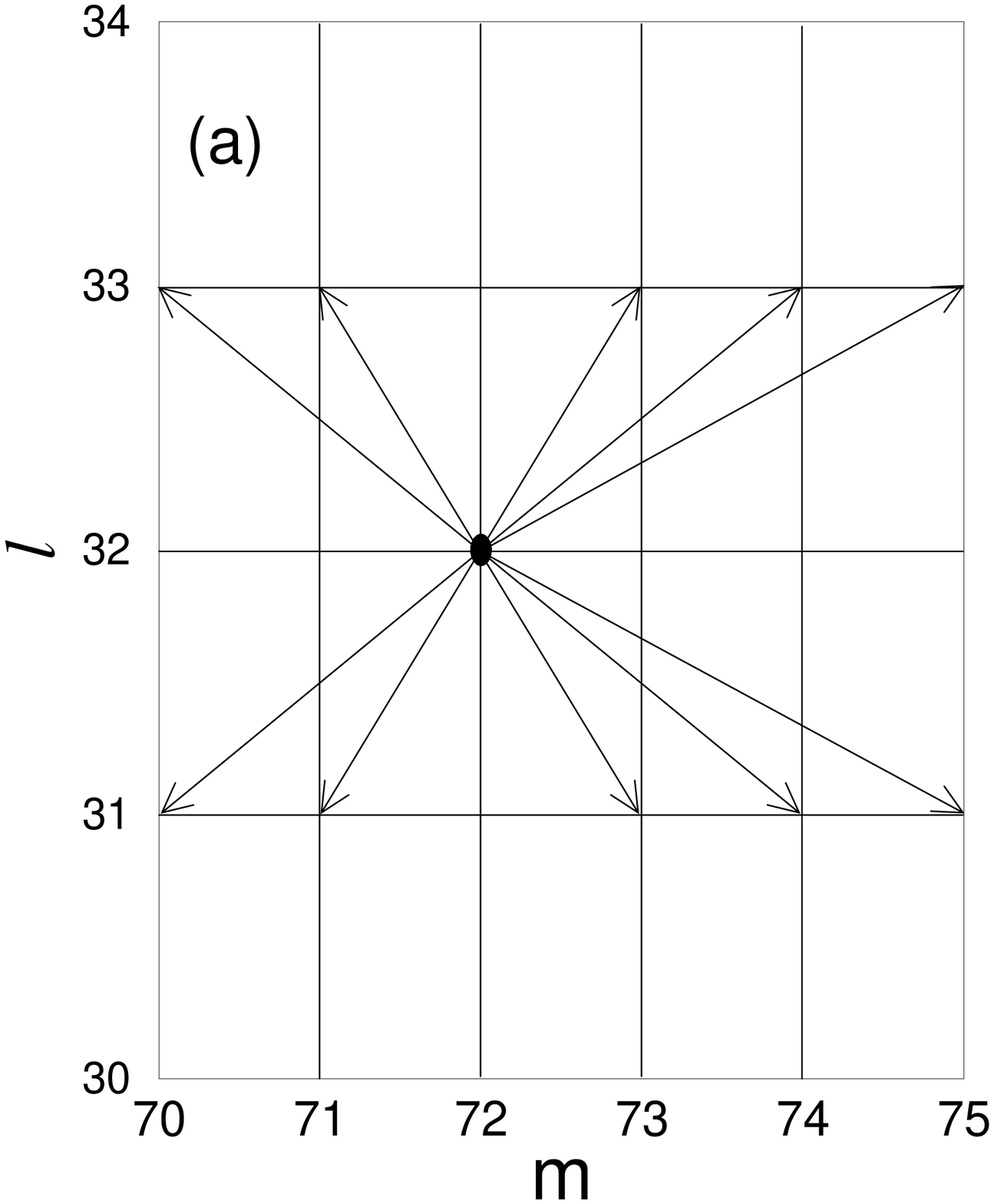,width=7cm,height=7cm}
       \psfig{file=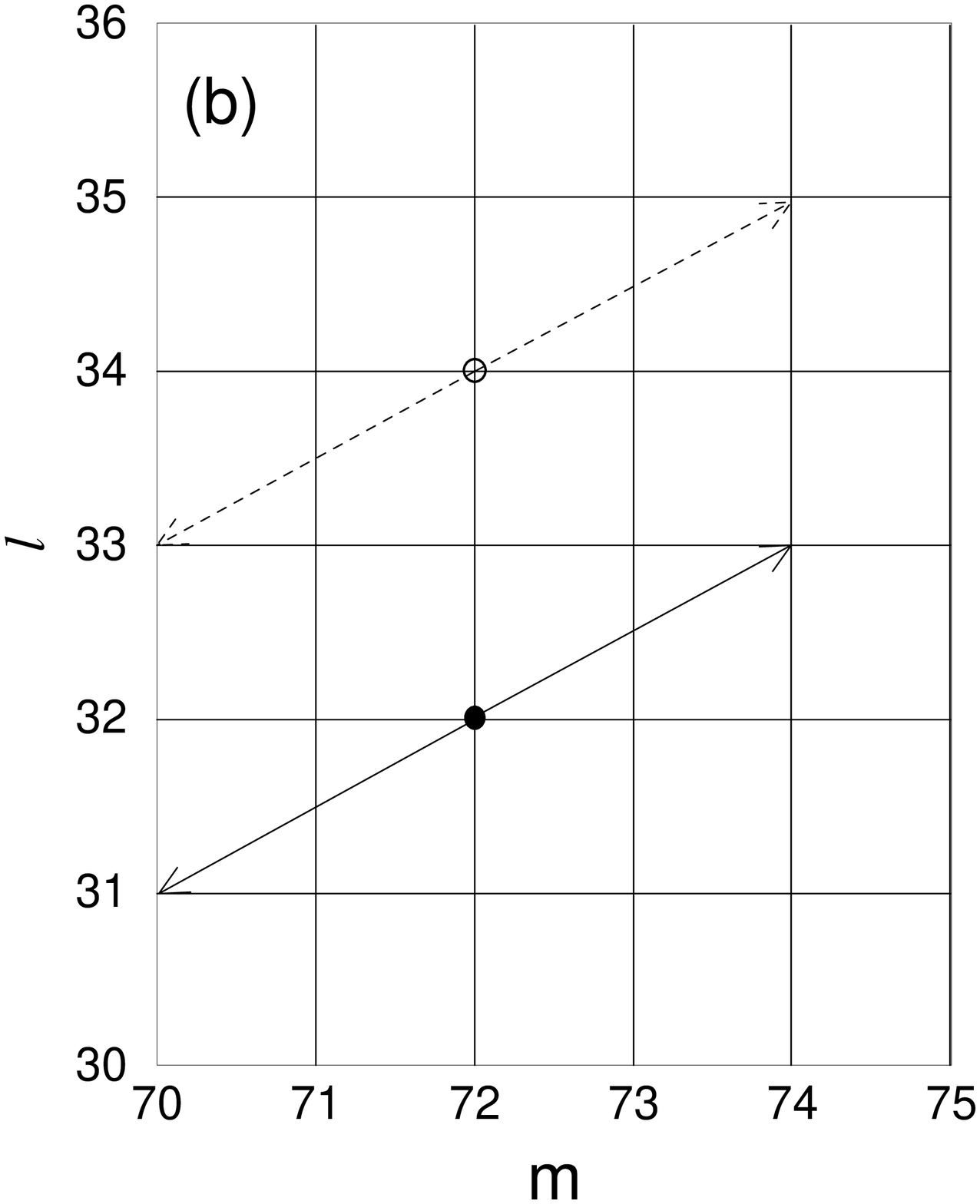,width=7cm,height=7cm}}
 \vspace{0.2cm}
 \caption{(a) Some possible transitions on a 2D
 lattice given by Eq.~(\ref{2DD}). (b) Possible transitions on a 2D
 lattice at small $\epsilon$, given by Eq.~(\ref{al_eq}). $N=2$.}
 \end{center}
 \label{fig:1}
 \end{figure}

Substitution (\ref{13}) in (\ref{11}) gives the equations for  $A_m^{q}$,
\begin{equation}
\label{15}
 \sigma_q A_m^{q}(\varphi)=-ih\mu{{dA_m^{q}(\varphi)}\over{d\varphi}}+
 h(m+1/2)A_m^{q}(\varphi)+
\end{equation}
$$
{{\epsilon}\over{2}}\sum_{n=-m}^\infty\Bigg(e^{-i\varphi}F_{m,m+n}
+e^{i\varphi}F^*_{m,m+n}\Bigg)
A_{m+n}^{q}(\varphi),\quad \varphi\equiv\mu\tau.
$$
Expanding the function $A_m^{q}(\varphi)$ in a Fourier series,
\begin{equation}
\label{16}
A_m^{q}(\varphi)=\sum_{l=-\infty}^\infty
A^{q}_{m,l}e^{-il\varphi},
\end{equation}
we derive the following equation for the amplitudes, $A^{q}_{m,l}$,
\begin{equation}
\label{2DD}
E_q A^{q}_{m,l}=h(m-\mu l)A^{q}_{m,l}+{{\epsilon}\over{2}}
\sum_{n=-m}^\infty\Bigg
(F_{m,m+n}A^{q}_{m+n,l+1}+F^*_{m,m+n}A^{q}_{m+n,l-1}\Bigg),
\end{equation}
where $E_q=\sigma_q-h/2$.
Equations (\ref{2DD}) can be interpreted as a problem 
for electron localization on a 2D lattice.
Indeed, one can consider the complex coefficients, 
$A^{q}_{m,l}$, as the
complex amplitudes of probability to find an electron on a 2D lattice at the
site $(m,l)$, where $0\le m<\infty$, $-\infty<l<\infty$.
Some possible transitions in the system (\ref{2DD}) are shown in 
Fig.~1~(a). The particle, initially located at the site with the indexes
$(m_0,l_0)$ can jump at the sites $(m_0\pm n,l_0\pm 1)$, where $n$ is
an integer number.   

\section{The resonance approximation}
When the interaction amplitude is small, 
$\epsilon\ll 1$, the 2D SSS described by Eq.
(\ref{2DD}) can be reduced to the 1D SSS described by the equation with only one index.
We divide both parts of Eq.~(\ref{2DD}) on $\mu$
and, taking into account that $\delta\ll N$ and $1/\mu\approx 1/N-\delta/N^2$, obtain
\begin{equation}
\label{2DDD}
{E_q\over\mu} A^{q}_{m,l}=
h\left(\frac mN-{\delta m\over N^2}-l\right)A^{q}_{m,l}+
{{\epsilon}\over{2\mu}}\sum_{n=-m}^\infty
\Bigg(F_{m,m+n}A^{q}_{m+n,l+1}+F^*_{m,m+n}A^{q}_{m+n,l-1}\Bigg).
\end{equation}

We assume that  $\delta$ is small, so that $\delta m/N^2 \ll 1$
for all considered values of $m$, or $\delta=0$.
Then, in the zeroth order approximation we have from Eq.~(\ref{2DDD}),
\begin{equation}
\label{zero_order1}
\left(\frac mN-l\right)A^{q}_{m,l}=\frac{E_q^{(0)}}{\mu h} A^{q}_{m,l}.
\end{equation}
It follows from Eq.~(\ref{zero_order1})
that if $A^{q}_{m,l}\not=0$, 
than $E_q^{(0)}/\mu h=(m/N)-l$. Since the ratio, $E_q^{(0)}/\mu h$, is defined
by modulus 1~(see Eq.~(\ref{U})), we can write, $E_q^{(0)}/\mu h=0$
\footnote{We assume that $(m/N)-l$ is an integer
for some initial state $m_0$.
If not, one may introduce the quasienergy, $E_q'=E_q-\mu h\{m_0/N\}$
and a new index, $m'=m-\{m_0/N\}N$, and solve
Eqs.~(\ref{2DDD})~and~(\ref{zero_order1})
for $E_q'$ and $m'$ instead of $E_q$ and $m$.
Here $\{x\}$ is the fractional part of $x$.}
Then Eq.~(\ref{zero_order1}) takes the form,
\begin{equation}
\label{zero_order}
(m-Nl)A^{q}_{m,l}=0.
\end{equation}
Hence,
\begin{equation}
\label{100}
A^{q}_{m,l}=0\qquad {\rm for}\qquad m\ne Nl,
\end{equation}
\begin{equation}
\label{101}
A^{q}_{m,l}\equiv A^{q}_{m}\qquad {\rm for}\qquad m=Nl.
\end{equation}

The next order approximation for $m=Nl$ yields,
\begin{equation}
\label{al_eq}
(E_{q}-h\delta m/N)A^{q}_{m}=\frac\epsilon 2
(F_{m,m+N}A_{m+N}^q+F^*_{m,m-N}A_{m-N}^q),
\end{equation}
where $E_q=E_q^{(1)}$ (we do not consider the higher order
approximations).
As one can see from Eq.~(\ref{zero_order}),
in the $m$-direction only hops on
the distance $N$ are allowed.   
Thus, the 2D problem, given by Eq.~(\ref{2DD}),
is reduced in the case $\epsilon\ll 1$ to the 1D problem described by
Eq.~(\ref{al_eq}).

The localization properties of the quantum states in the resonance
approximation, given by
Eq.~(\ref{al_eq}), are defined by the structure of the matrix
elements, $F_{m,m+N}$. If the matrix elements are periodic functions of
$m$, all the eigenstates
are extended and the spectrum is continuous.
This situation is common for solid-state systems~\cite{Harper}.
In the system under consideration, the matrix elements (\ref{m_el})
are non-periodic. On this reason, as
will be shown below (see also Refs.~\cite{31,21})
the quantum states are localized and the spectrum is discrete.

The matrix elements given by Eq.~(\ref{m_el}),
oscillate as a function of $m$.
At the points, $m_0$, where the matrix elements are close to zero, 
\begin{equation}
\label{Bessel}
F_{m_0,m_0+N}\sim J_N\left(\sqrt{2m_0h}\right)\approx 0, 
\end{equation}
the transition probability is very small. 
As a consequence, such points becomes the dynamical barriers to the 
probability flow~\cite{21}, and divide the Hilbert space, labeled by
index, $m$, into the 
relatively independent parts --- resonance cells~\cite{11}.
Most of the eigenstates given by Eq.~(\ref{al_eq}) are
concentrated inside these cells. The average localization length,
$\langle\lambda_i\rangle$, for the states in the $i$-th cell does
not exceed the size of the cell. The cell boundaries are defined by 
Eq.~(\ref{Bessel}), i.e. $\left\langle\lambda_i\right\rangle \le m_{i+1}-m_i$, where
$m_i$ and $m_{i+1}$ satisfy Eq.(\ref{Bessel}), so that  
$\sqrt{2m_ih}$ and $\sqrt{2m_{i+1}h}$ are, respectively, the 
$i$-th and $i+1$-th roots of the Bessel function in (\ref{Bessel}).

Some characteristic QE functions given by
Eq.~(\ref{al_eq}) are illustrated in Figs.~2~(b)~-~(e), for small $\epsilon$.
The boundaries of the resonance cells are
marked by arrows. One can see from Figs.~2~(b)~-~(e), that the eigenfunctions
are localized inside the cells, but, on the other hand,
each  eigenfunction is delocalized over $m$ inside a single cell.
For example, for the initial states in Fig.~1~(b) with $m=72$
the transitions will occur in the region $66<m<176$ 
(inside the first cell in Fig.~2~(a)).
Note, that for small values of $\epsilon$ and when $\delta=0$, the
localization properties of our system are independent of $\epsilon$.
This means that an arbitrary small perturbation, $\epsilon$, initiates 
transitions between the sites on the effective rectangular lattice.
For small values of $\epsilon$, these transitions take place on
1D sub-lattices shown in Fig.~1~(b).

 \begin{figure}[tb]
 \centerline{\psfig{file=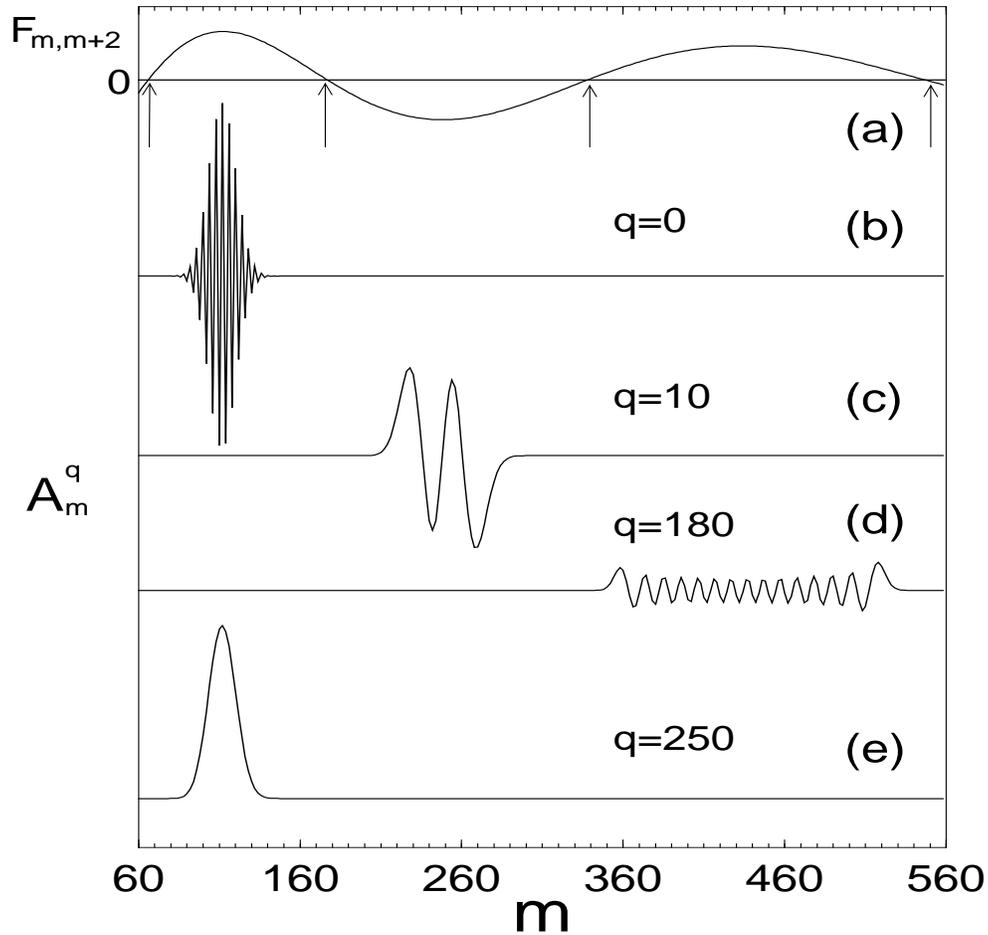,width=13cm,height=13cm}}
 \caption{(a) The matrix elements (in arbitrary units),
 $F_{m,m+2}$, and (b)-(e) some
 characteristic QE functions given by Eq.~(\ref{al_eq}) with
 $h=0.2$, $\epsilon=0.02$, $\delta=0$, $N=2$.
 Only even values of $m$ are included.}
 \label{fig:2}
 \end{figure}

Except for the localized (in the resonance cells) eigenfunctions
there exists few eigenfunctions which are delocalized over
several resonant cells (DF).
One of these representative eigenfunctions is shown in Fig.~3. The
DFs have maxima in the regions near the boundaries of the cells marked 
in Fig.~3 by arrows. Thus, if the initial state is located
near the boundary of a resonance cell, say, at $m=66$
in Fig.~3 (in the region near the first arrow), then this state will
propagate at a large distance in $m$, over the 1D sub-lattice.
This distance can be much larger than the size of 
the single resonance cell~\cite{31}.

 \begin{figure}[tb]
 \centerline{\psfig{file=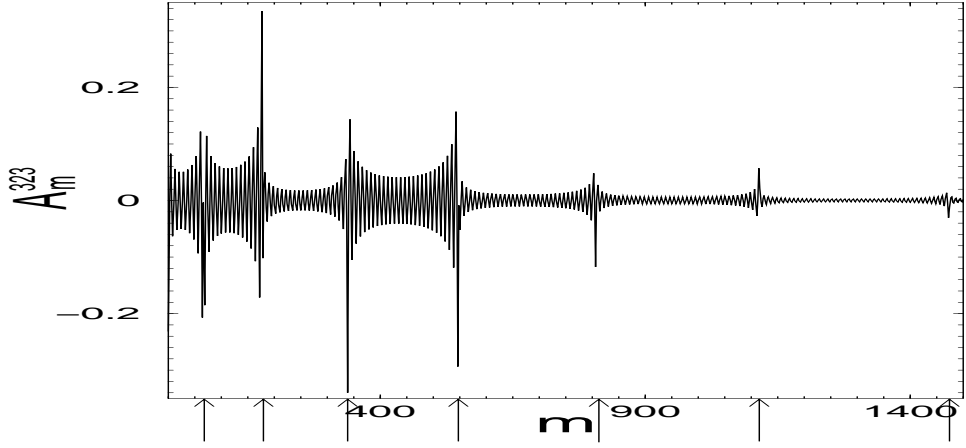,width=13cm,height=6cm}}
 \vspace{0.2cm}
 \caption{The characteristic DF
 (at even $m$), $q=323$, $N=2$, $h=0.2$, $\epsilon=0.02$, $\delta=0$.}
 \label{fig:3}
 \end{figure}

 \begin{figure}[tb]
 \mbox{\psfig{file=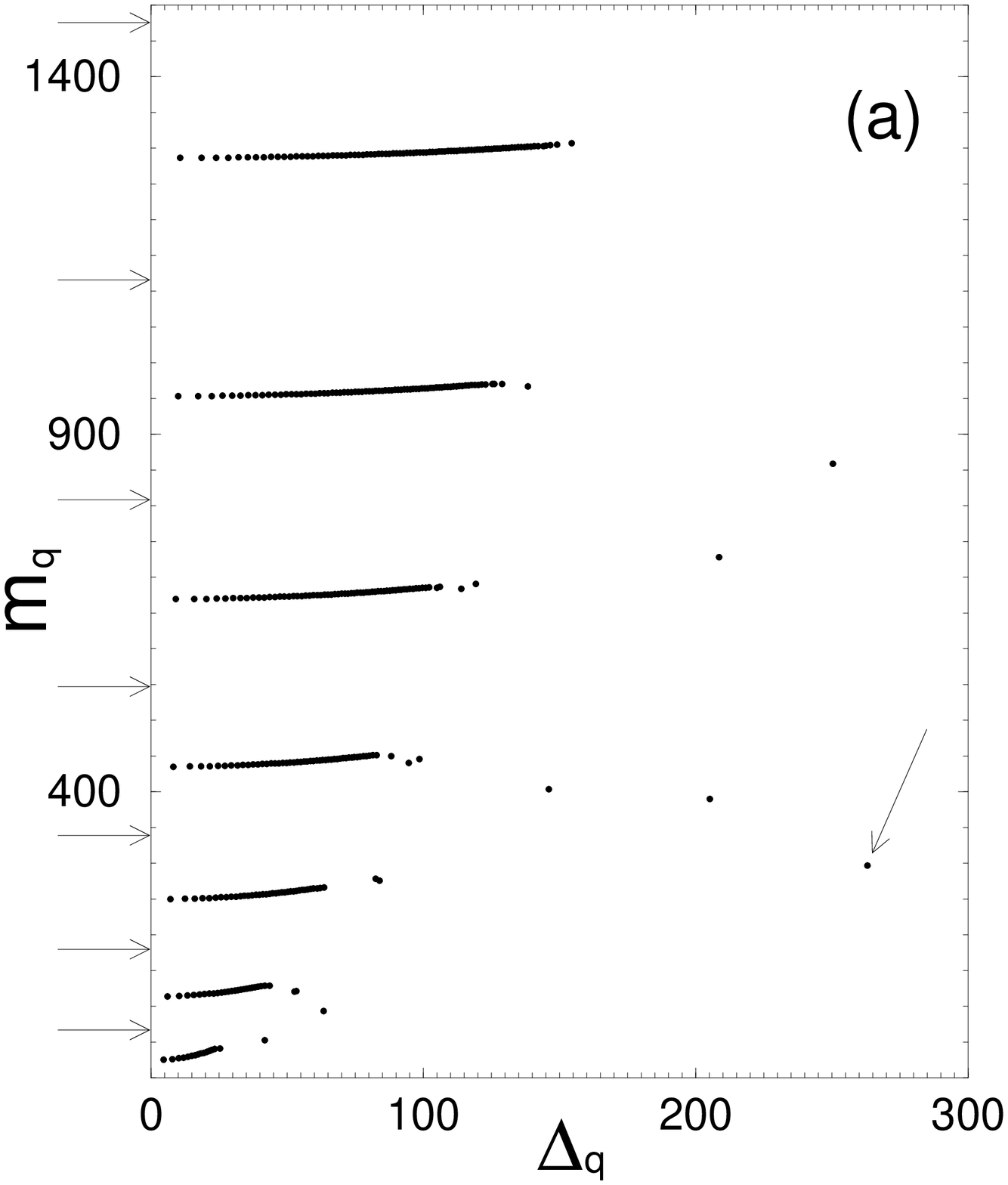,width=8cm,height=8cm}
       \psfig{file=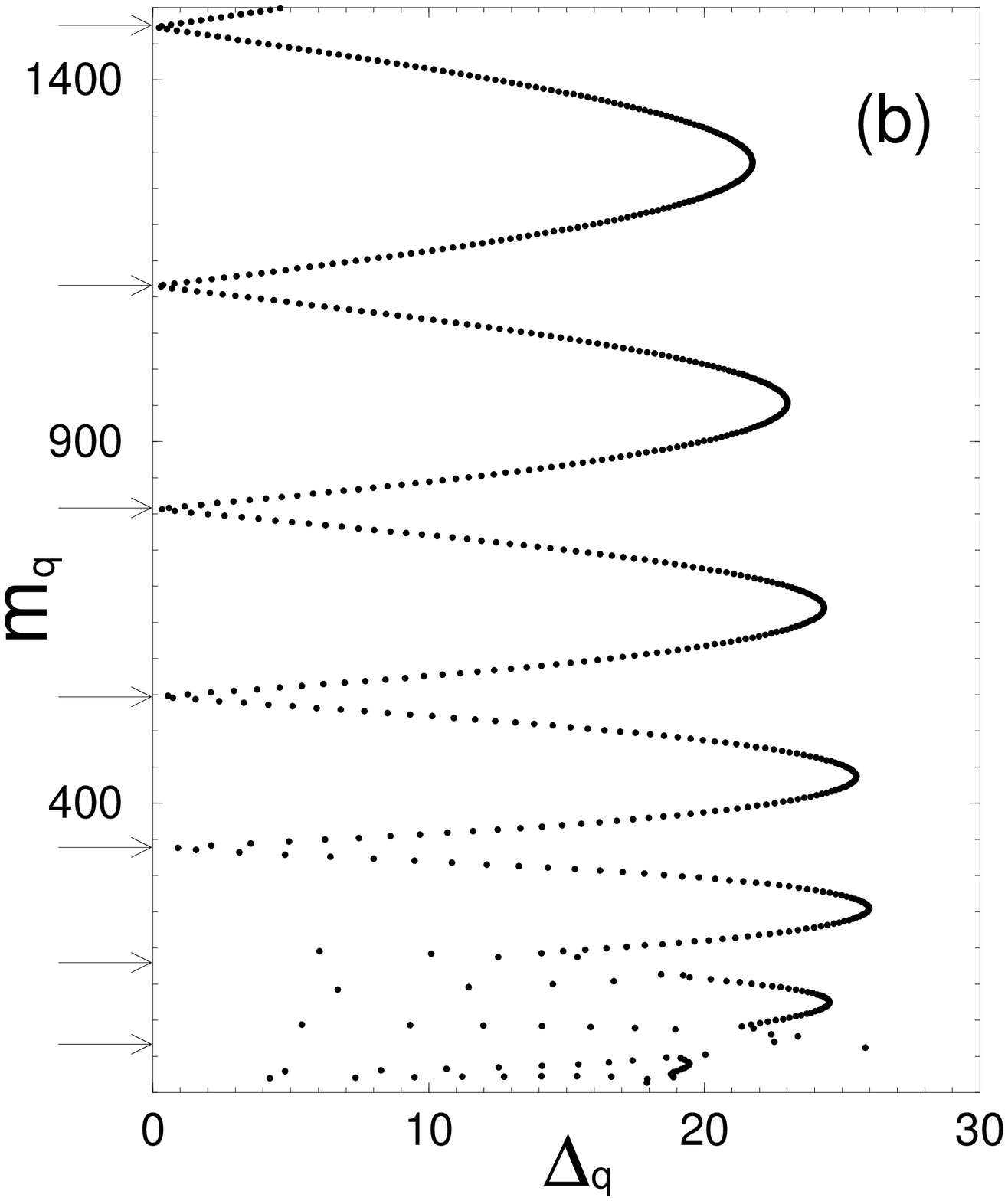,width=8cm,height=8cm}}
 \caption{The plot $m_q$ versus $\Delta_q$ for $N=2$, $h=0.2$,
 $\epsilon=0.02$ and (a) $\delta=0$, (b)
 $\delta=0.001$. The boundaries of the resonance cells are marked by
 the arrows on the $m_q$-axis.}
 \label{fig:4}
 \end{figure}

The structure of the eigenfunctions
can be better understood from the plot of the mean,
$m_q=\sum_m |A_m^q|^2 m$, versus variance, 
$\Delta_q=[\sum_m |A_m^q|^2(m-m_q)^2]^{1/2}$,
presented in Fig.~4~(a).
Each eigenfunction, $A_m^q$, is represented by one point in the figure.   
One can see that most of the eigenfunctions are localized inside the
resonance
cells since their means are located inside the cells and their variances
do not exceed the size of the cell. Each row on the figure is formed by 
the eigenfunctions of one cell. If the initial state is located inside the 
resonance cell, 
the eigenfunctions of this particular cell 
define the quantum dynamics. These states make the quantum dynamics localized inside the cell and, at the same time,
delocalized over the states, $m$, inside the cell~\cite{21}.
In the corresponding solid-state model (\ref{al_eq}), the localization length
at small $\epsilon$ can be identified with the size of the resonance cell. 

The DFs are represented in Fig.~4~(a)
by the scattered points with large variances. 
One of the DFs, marked in Fig.~4~(a) by an arrow, is shown in Fig.~3.
The DFs can not be attributed to a definite resonance cell
since their variances are larger than the size of single cells. 
As a consequence, the DFs cause delocalization of the states initially 
concentrated near the boundaries of the cells. However, as 
shown in Ref.~\cite{21}, the localization length remain finite,
because the matrix elements (\ref{m_el}) are non-periodic and 
their amplitudes decrease with $m$ increasing  
(as $m^{-1/4}$ at $m\gg 1$). 
    
In the case when the detuning from the resonance
(see (\ref{1})) is not equal to zero ($\delta\ne 0$),
the character of localization depends on the position of an initial state, 
$m_0$ (see Fig.~4~(b)). In the region
$m_0\gg m_{max}=\epsilon N/h\delta$ all the 
states remain exponentially localized in $m$, since in this case
Eq.~(\ref{al_eq}) has the solution,
\begin{equation}
\label{delta_solution}
E_q=(h\delta m/N)\delta_{m,q},\qquad A_{m}^q=\delta_{m,q}.    
\end{equation}
If $m_0\ll m_{max}$ the above discussed effect of localization
over the resonance cells takes place.   
In the intermediate case, when 
$m_0\ge m_{max}$, the character of localization depends on the position 
of the state $m_0$ inside the resonance cell.  (For the parameters in 
Fig.~4~(b) $m_{max}=200$.)
If $m_0$ is located 
near the boundary of the resonance cell where the condition
(\ref{Bessel}) is satisfied, Eq.~(\ref{al_eq}) has the localized
solution (\ref{delta_solution}). 
Most delocalized functions
have their mean, $m_q$, at the center of a resonance cell.

As follows from Fig.~4~(b), in the region $m_0\ge m_{max}$ the DFs
are absent, since the variance of each function
is much less than the size of the cell, whose boundaries are marked 
in Fig.~4~(b) by arrows. Moreover, the variances of the eigenstates
in the near-resonance case in Fig.~4~(b) are substantially smaller than the
variances in the exact resonance case shown in Fig.~4~(a). Hence, at small $\epsilon$, the
increase of the value of the detuning, $\delta$, always leads 
to localization of the quantum 
states in the discussed model.

Most of the localization properties of the eigenfunctions, given by 
Eq.~(\ref{al_eq}), are the quantum manifestation of the classical
behavior in the phase space. The classical phase space
in the variables $(kr(I),\,\theta)$), where
$\theta=N\vartheta$, mod $2\pi$,
generated
by the exact classical Hamiltonian (\ref{H_cl}), is shown in Fig.~5~(a)
for the exact resonance case ($\delta=0$)
and in Fig.~5~(b) for the near resonance case ($\delta=0.001$).

 \begin{figure}[tb]
 \mbox{\psfig{file=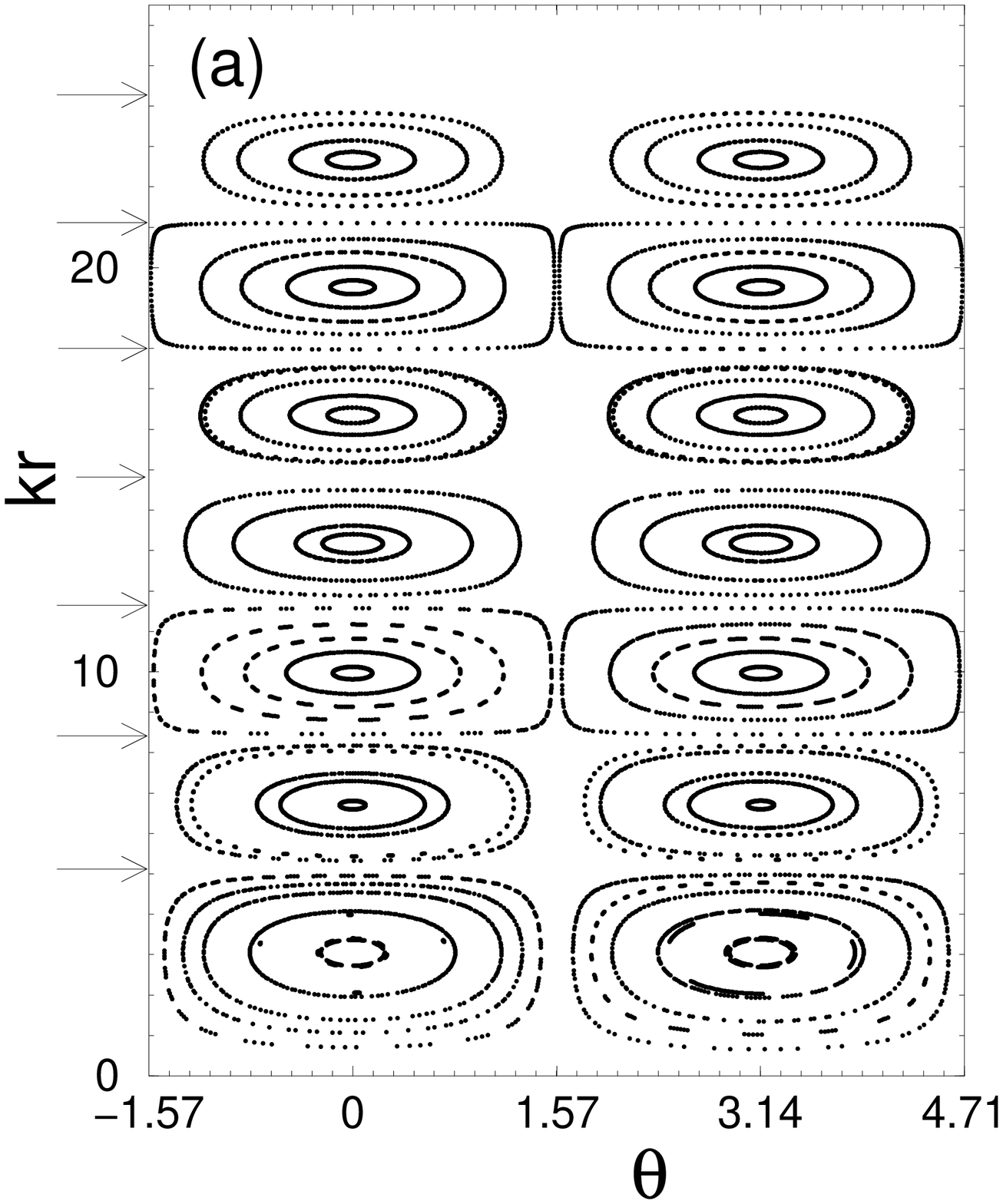,width=8cm,height=8cm}
       \psfig{file=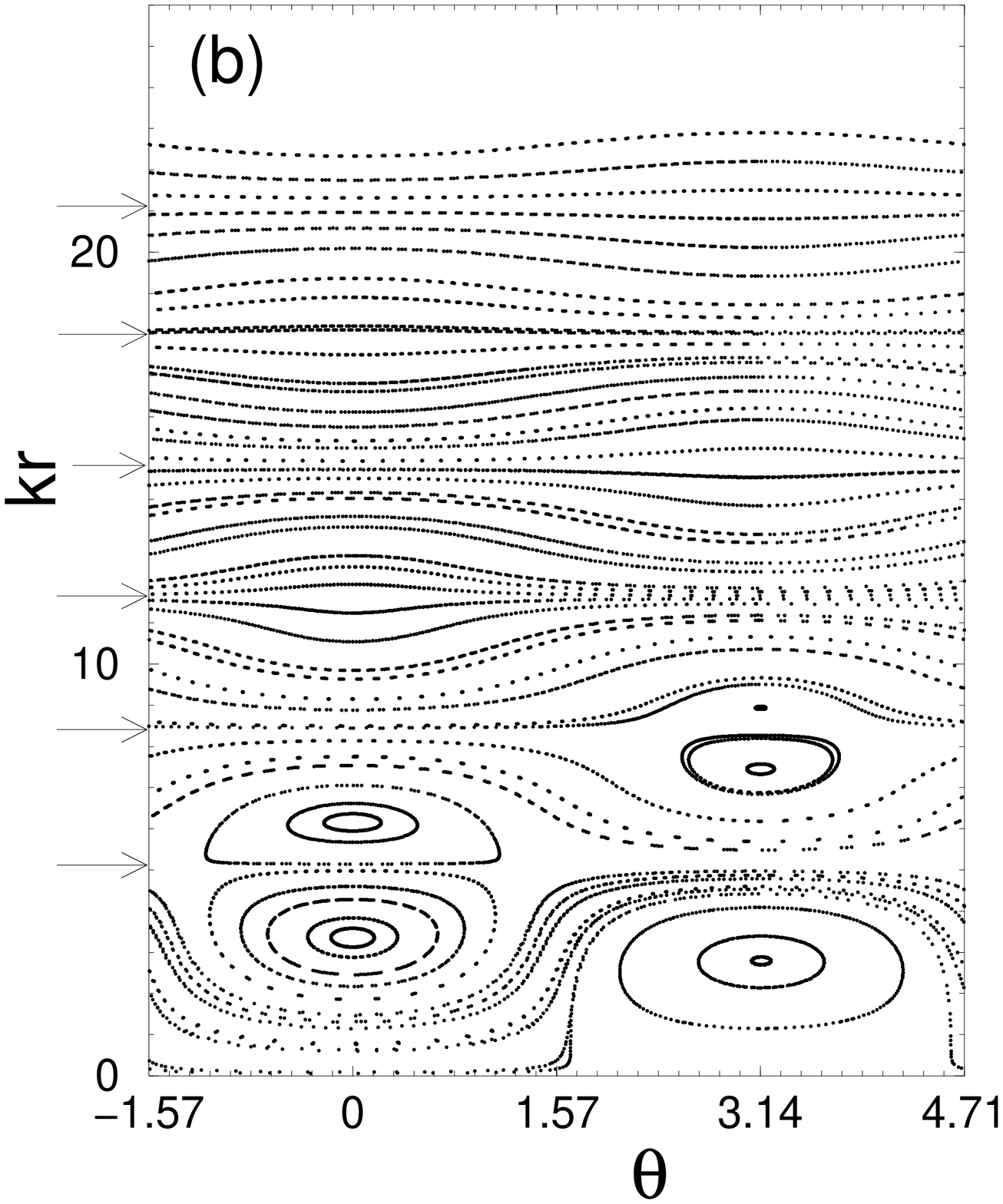,width=8cm,height=8cm}}
 \caption{The classical phase space for (a) $\delta=0$, (b) $\delta=0.001$; 
 $N=2$, $\epsilon=0.02$. The boundaries of the resonance cells are marked by 
 arrows.}
 \label{fig:5}
 \end{figure}

As one can see from Fig.~5~(a), in the case $\delta=0$ the
classical phase space is divided into the resonance cells.
(Fig.~5~(a) shows only the first seven cells.)
The boundaries of the cells, $kr(I_i)=\sqrt{2I_i}$, are marked 
in Figs.~5~(a),~(b) by arrows.
As shown in Refs.~\cite{11,tobe}, in the quasiclassical limit
the $i$-th boundary of the classical cell, $kr_i$, in Figs.~5~(a),~(b)
corresponds to $i$-th boundary of the quantum cell, $m_i$, 
in $m_q$ - axis in Figs.~4~(a),~(b), so that
$kr_i=\sqrt{2I_i}=\sqrt{2hm_i}$. 

Each row of points in Fig.~4~(a) is formed by the eigenstates
responsible for the dynamics in the corresponding quantum cell. 
From comparison with the classical dynamics in
phase space we can now describe
the localization properties of the quantum states, discussed above.  
Each value of $m$ in the quantum system corresponds to a 
quantized classical action, $I_m=mh$,
or to the quantized dimensionless oscillation amplitude, $kr_m=\sqrt{2mh}$.
Each value of action $I_m$ (or $kr_m$) corresponds to the
set of classical trajectories. Moving along some classical trajectory
the particle with some initial value of action, $I_m$,
can accept the other values in the
interval $I_{m_1}<I_m<I_{m_2}$. The corresponding
eigenstate will be delocalized over the unperturbed states with the 
numbers $m$ in the interval, $m_1<m<m_2$.
From the form of the trajectories in Fig.~5~(a) one can see that
in the case of exact resonance
all quantum states of the single quantum cell should be delocalized
over the resonance cell, since in the phase space both the extremal
values, $I_{m_1}$ and $I_{m_2}$, which limit the resonance cell,
can belong to the same trajectory.

Similar arguments can be used to analyze the quantum-classical
correspondence in the near resonance case.
As follows from Fig.~5~(b), at $\delta\ne 0$
in the phase space there is only a finite number of resonance cells
(two cells in Fig.~5~(b)). Thus, there is a finite number of quantum resonance cells in the
Hilbert space in Fig.~4~(b) (the first two cells). In the
off-resonant region (3th-7th cells in Fig.~4~(b))
the degree of delocalization of eigenstates
depends on the position of the state
in the cell destroyed by the finite detuning, $\delta$.
In Fig.~5~(b) the least curved trajectories are located
near the separatrices, while the the most curved trajectories are 
located near the centers of the destroyed cells. As a consequence, 
in the quantum model the eigenfunctions in Fig.~5~(b) have smallest
variance in the region near 
the separatrices and largest variance near the centers of the 
destroyed cells.

\section{The localization length in the case of strong interaction}
In the previous sections we considered the dynamics only at small  
perturbation amplitude, $\epsilon\ll 1$. At large values of
$\epsilon$ the dynamical chaos appears in the classical~\cite{Z}
and quantum~\cite{31,101} MPO. As will be shown below, all
quantum states in the chaotic area are delocalized over 
the whole chaotic region.  
The chaotic dynamics in MPO corresponds to hops in different directions 
in the SSS, as shown in Fig.~1. By estimating the size of
the chaotic motion in the MPO we will estimate below the localization length 
in the SSS. 

 \begin{figure}[tb]
 \centerline{\psfig{file=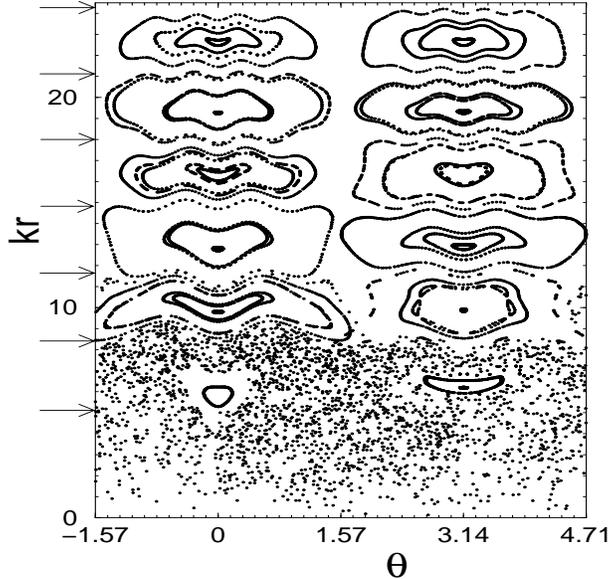,width=8cm,height=8cm}}
 \caption{The classical phase space in the case $\epsilon=3$.
 Other parameters are: $\delta=0$, $N=2$.
 The boundaries of the resonance cells are marked by arrows.}
 \label{fig:6}
 \end{figure}

 \begin{figure}[tb]
 \mbox{\psfig{file=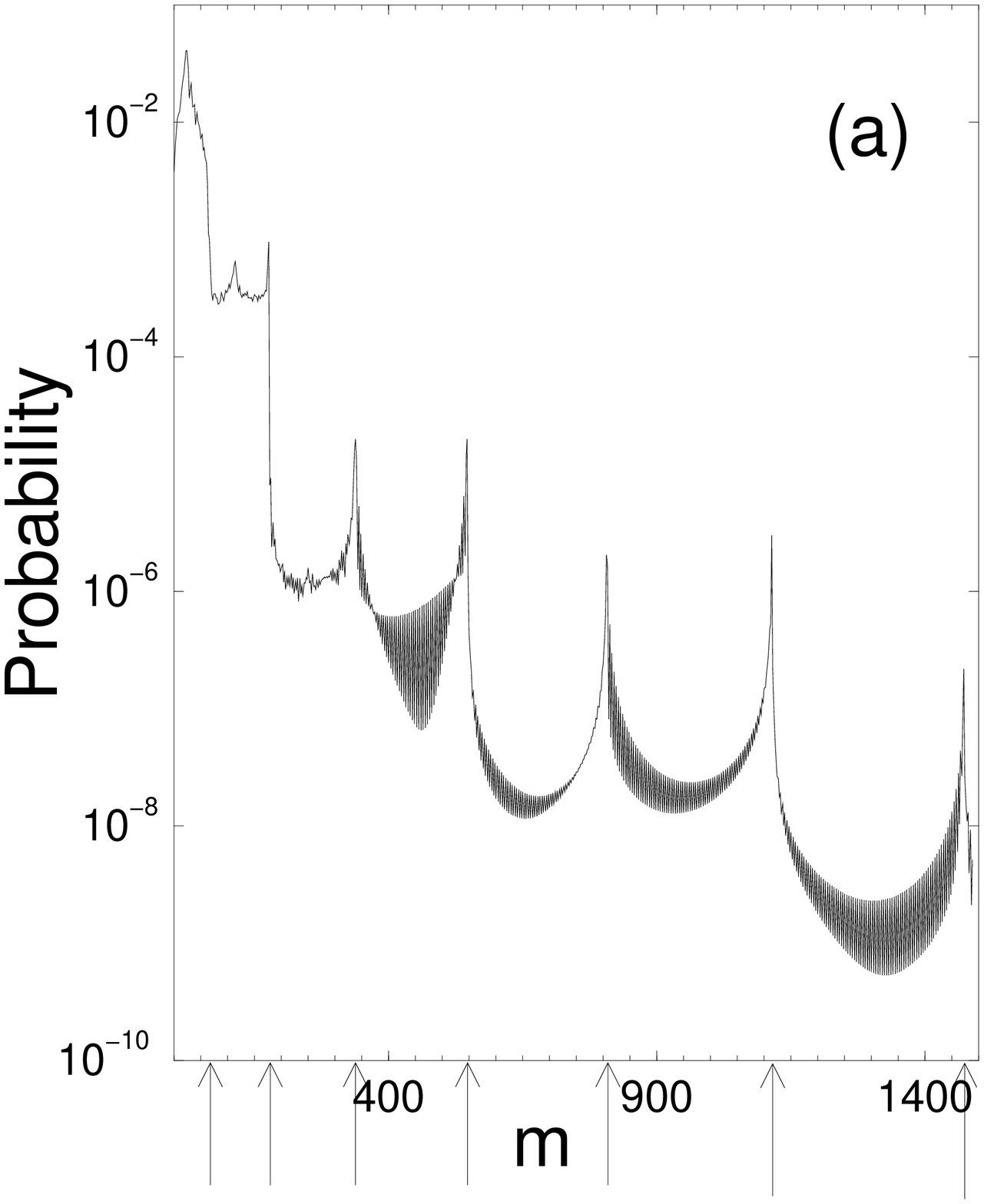,width=7.5cm,height=7.5cm}\hspace{3mm}
       \psfig{file=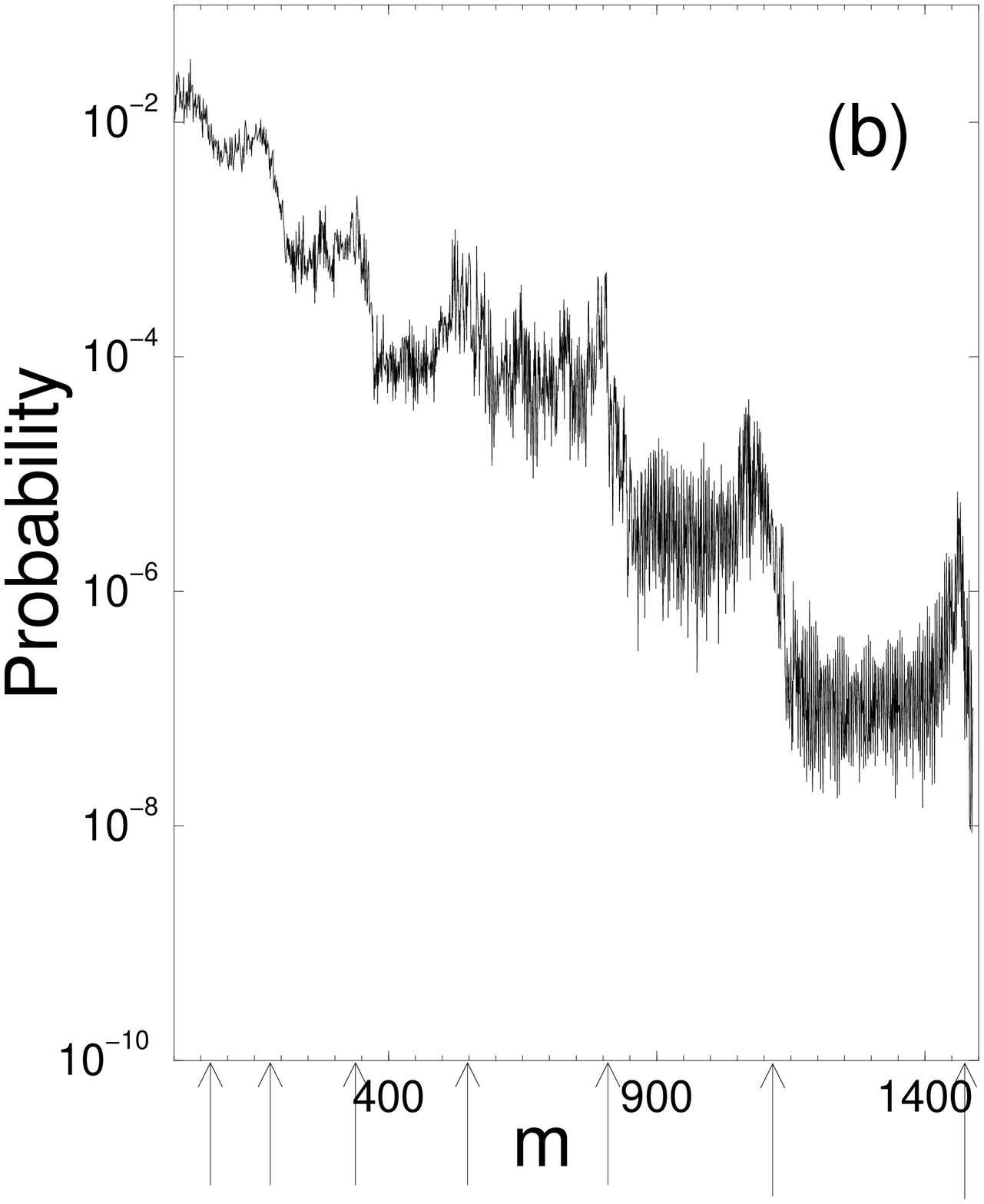,width=7.5cm,height=7.5cm}}
 \caption{The time-averaged probability distribution.
 (a) $\epsilon=0.02$, the averaging has been performed over 100
 realizations in the time-interval $\tau=5000-105000$
 (only the probability at even $m$ is shown);
 (b) $\epsilon=3$, where the averaging has been performed over 100
 realizations in the time-interval $\tau=500-10500$.
 Other parameters are: $h=0.2$, $\delta=0$, $N=2$.
 The boundaries of the resonance cells are marked by arrows.}
 \label{fig:7}
 \end{figure}

In Fig.~6 the classical phase space is shown for $\epsilon=3$. One can see
that in the first two cells the motion is mainly chaotic while
in other cells the motion remains mainly regular.
The quantum probability distribution is illustrated in Fig.~7~(a)
for the case of small $\epsilon$ and in Fig.~7~(b) for the case
$\epsilon=3$.
The initial state was taken in the form $c_m(0)=\delta_{m,m_0}$
with $m_0=30$ (in the center of the first cell in Figs.~7~(a),~(b)).
As follows from Fig.~7~(a), the quantum particle can
tunnel (see also Ref. \cite{21}) from the initial (first) cell to other resonance
cells unlike the classical case, where practically all the trajectories
in the phase space are confined inside the resonant cells (see Fig.~5~(a)).

\vspace{3mm}
 \begin{figure}[tb]
 \mbox{\psfig{file=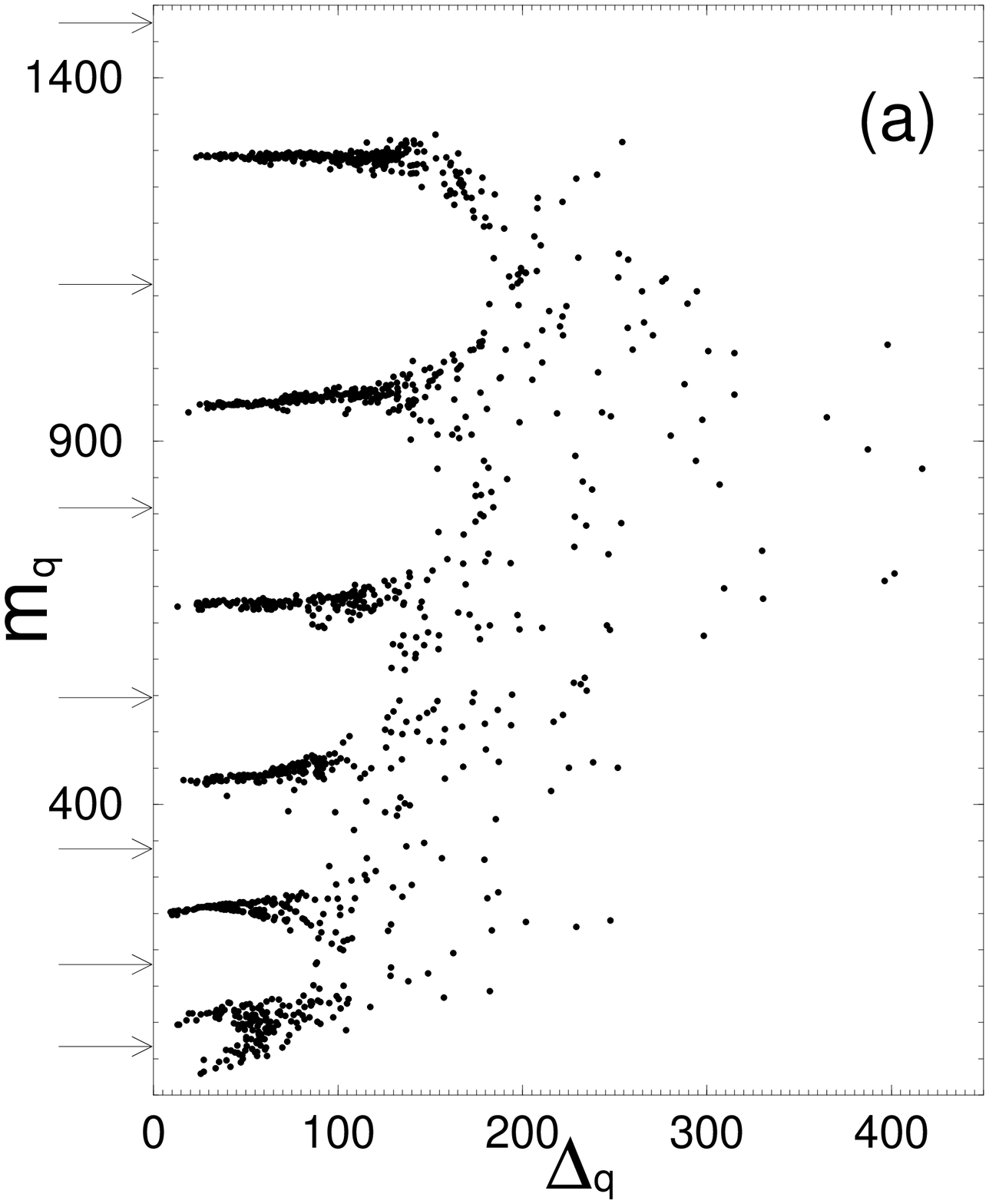,width=7.5cm,height=7.5cm}
       \psfig{file=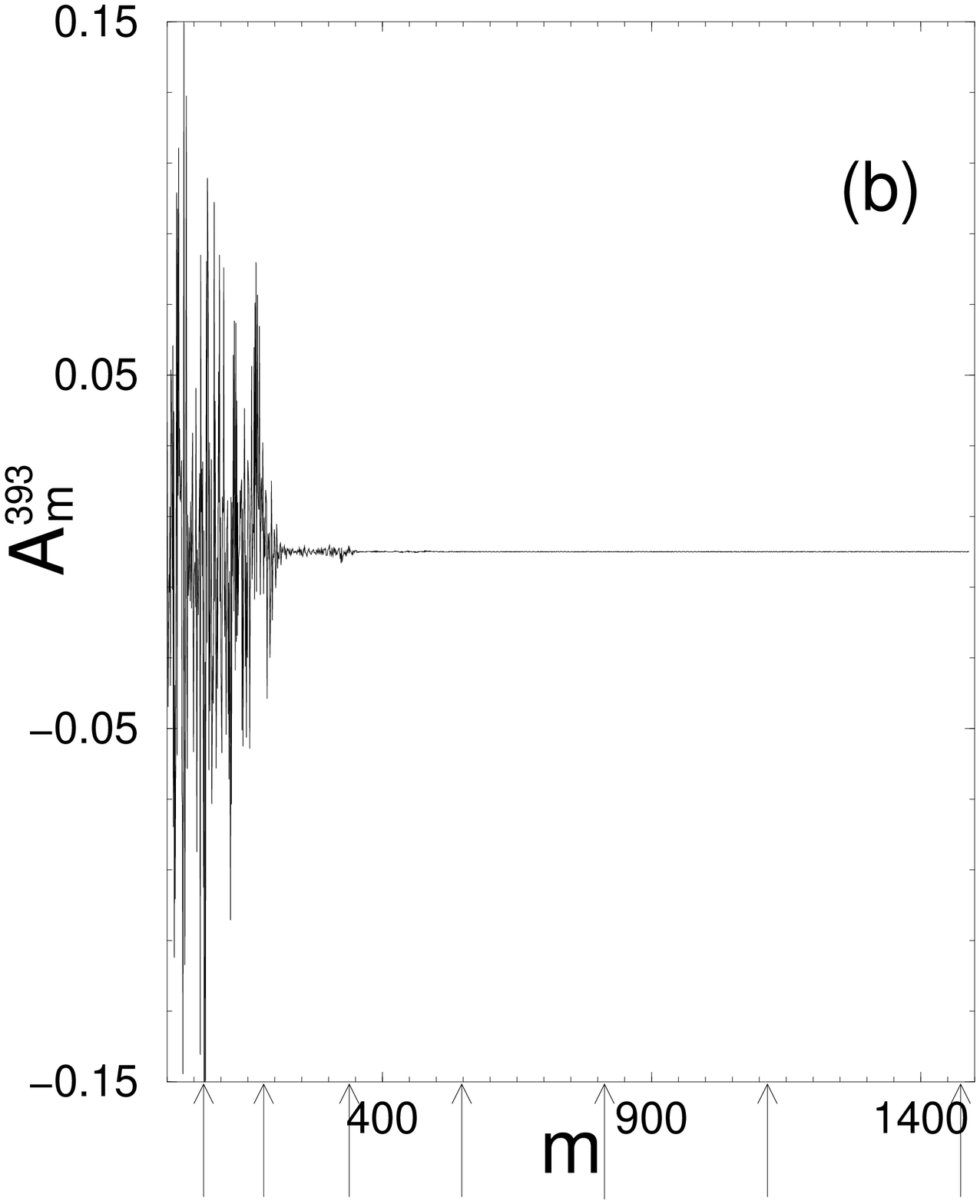,width=7.5cm,height=7.5cm}}
 \caption{(a) The plot $m_q$ versus $\sigma_q$
 and (b) the characteristic QE eigenfunction located in the
 chaotic region for $\epsilon=3$ and $h=0.2$, $\delta=0$, $N=2$.
 The boundaries of the resonance cells are marked by arrows.}
 \label{fig:8}
 \end{figure}

When $\epsilon$ increases, the probability distribution
in Fig.~7~(b) increases
in most of the quantum cells, which corresponds to 
chaotization of motion in the classical phase space.
In Fig.~8~(a) we show
the plot $m_q(\sigma_q)$ for the case $\epsilon=3$, and in Fig.~8~(b)
the characteristic QE function located in the
chaotic area is illustrated. 
As one can see from Fig.~8~(a), almost all QE states in the area
of the first two cells are delocalized over both the cells.
In other words, the QE states are 
localized inside the chaotic area (first two cells),
but not inside the single cells, as in the case of small $\epsilon$
in Fig.~4~(a). In the chaotic regime,
one can find a quantum particle with equal probability 
in any unperturbed state, $m$, inside the chaotic sea, independently
of the position and the form of an initial state located in this region
(see first two cells in Fig.~7~(b) and 8~(a),~(b)).
When $\epsilon$ increases, more classical and quantum cells become chaotic.   
This results in increasing the area of delocalization of the quantum chaotic
states. Thus, in the regime of chaos 
the localization length in the SSS may be
identified with the size of the chaotic area in the TPS.

\section{Conclusion}
The resonance approximation, given by Eqs.~(\ref{zero_order})~-~(\ref{al_eq})
can be interpreted in the following way.
We can re-write the Hamiltonian matrix in Eq.~(\ref{2DD}) in the form,
\begin{equation}
\label{H_matr}
H_{m,l;m+n,l'}=H^0_{m,l;m+n,l'}\delta_{n,0}\delta_{l,l'}
+F_{n,n+m}\delta_{l',l+1}+F_{n,n+m}^*\delta_{l',l-1}.
\end{equation}
The unperturbed motion, given by the zeroth order Hamiltonian,
$H^0_{n,l}=h(n-\mu l)$, takes place on the infinite 1D energy surface
(line) in the 2D space.
The character of the motion on the energy
surface depends on the form of the matrix elements of the interaction
potential, $\langle m,l|V(X,\tau)|m',l'\rangle=F_{m,l;m'l'}$.
If $V(X,\tau)$,
is linear in $X$,
for example, $V(X,\tau)=\epsilon X\cos\mu\tau$, and $\mu=N$, where
$N=1,\,2,\,\dots$, then the motion is unlimited~\cite{Feynman,Schwinger}.
When the interaction potential is nonlinear in $X$,
$V(X,\tau)=\epsilon \cos(X-\mu\tau)$ at $\mu=N$, or
when the period of the interaction potential, $\mu$, is not equal
to $N\times$ (a distance between the sites of the 2D cell), i.e. when
$\delta\ne 0$, then, as demonstrated in this paper
(see also Ref. \cite{21}), the quantum states
are localized.

It is necessary to note that, as was shown in~\cite{101},
the quantum and classical dynamics in the chaotic regime
is practically independent of the detuning, $\delta$,
when $\epsilon\gg \delta$.
So, the results concerning the chaotic dynamics also remain valid in the
near-resonance case.

In summary, the regular and chaotic classical and quantum dynamical
regimes are analyzed in the system of an ion trapped in a linear ion
trap and interacting with two laser field with close frequencies. This
system is modeled using
a quantum oscillator perturbed by a monochromatic wave (MPO). It is shown
that the problem of dynamical stability  in this system
corresponds to the problem of electron localization in a 2D solid-state
system (SSS). The resonance approximation is used to decrease the effective
dimensionality of the corresponding solid-state system. This can be done
in the case of relatively small interaction between the trapped ion and
the laser fields. In the SSS
this case corresponds to weak interaction between the sites of the
2D lattice. Increasing the interaction amplitude
results in delocalization
of the quantum states over the sites of the 2D cell. The area
of the delocalization in the SSS at strong interaction
in the chaotic area may be identified with the size
of the chaotic sea in the MPO. Our results  provide
understanding of the mechanism
of stability of an ion trapped in a linear ion trap. They also allow
one to estimate the characteristic dynamical regimes of the trapped ion
and to choose parameters required for dynamical stability.
 
\section{Acknowledgments}
We thank G. Chapline for useful discussions. The paper was supported by the Department of Energy under contract
W-7405-ENG-36 and by the National Security Agency.

\end{document}